\documentclass[pra, aps, twocolumn, floatfix, superscriptaddress, amsmath,  longbibliography]{revtex4-1}
\usepackage{graphicx}
\usepackage{graphics}
\usepackage{mathtools}
\usepackage{float}
\usepackage{amssymb}

\usepackage[dvipsnames]{xcolor}
\definecolor{darkblue}{rgb}{0.0, 0.0, 0.75}

\usepackage{color}
\usepackage[colorlinks=true,
            linkcolor=darkblue,
            urlcolor=darkblue,
            citecolor=darkblue]{hyperref}

	\usepackage[normalem]{ulem} 
	\definecolor{mgreen}{RGB}{1,123,0}

\def \bk{{\bf k}}

\def \br{{\bf r}}

\def \ms{\text{s}}

\def \mB{\mathrm{B}}
\def \ms{\mathrm{s}}

\def \mum{\mu\mathrm{m} }

\def \mms{\mathrm{ms}}

\newcommand{\uvec}[1]{\boldsymbol{\hat{\textbf{#1}}}}

\begin{document}
\title{Thermal suppression of demixing dynamics in a binary condensate}
\author{Vijay Pal Singh}
\affiliation{Quantum Research Centre, Technology Innovation Institute, Abu Dhabi, UAE}
\affiliation{Zentrum f\"ur Optische Quantentechnologien and Institut f\"ur Laserphysik, Universit\"at Hamburg, 22761 Hamburg, Germany}
\author{Luigi Amico}
\affiliation{Quantum Research Centre, Technology Innovation Institute, Abu Dhabi, UAE}
\affiliation{INFN-Sezione di Catania, Via S. Sofia 64, 95127 Catania, Italy}
\affiliation{Centre for Quantum Technologies, National University of Singapore 117543, Singapore}
%
\author{Ludwig Mathey}
\affiliation{Zentrum f\"ur Optische Quantentechnologien and Institut f\"ur Laserphysik, Universit\"at Hamburg, 22761 Hamburg, Germany}
\affiliation{The Hamburg Centre for Ultrafast Imaging, Luruper Chaussee 149, Hamburg 22761, Germany}
\date{\today}
%
%
\begin{abstract}
We investigate the demixing dynamics in a binary two-dimensional (2D) Bose superfluid using classical-field dynamics. 
By quenching the interspecies interaction parameter, we identify a strong and weak separation regime depending on the system temperature and the quench parameter. 
In the strong separation regime our results are in agreement with the inertial hydrodynamic domain growth law of binary fluids and a Porod scaling law for the structure factor at zero temperature is found. 
In the weak separation regime thermal fluctuations modify both the domain growth law and the Porod tail of the  structure factor. 
Near the superfluid transition temperature the scaling dynamics approaches the diffusive growth law of a 2D conserved field. 
We then analyze the demixing dynamics in a box cloud. 
For low quench we find distinctive domain dynamics dictated by the boundary condition. 
Otherwise, the dynamics are qualitatively similar to those of systems with periodic boundary conditions. 
\end{abstract}
\maketitle

\section{Introduction}
When two immiscible fluids such as water and oil are allowed to mix, they separate into two distinct phases \cite{onuki2002}.
Such phase separation phenomenon is very well established in science, with relevant implications for important technological applications \cite{Tateno2021, Wang2021, Mehta2022}. 
In physics, phase separation occurs in a  variety of condensed matter systems  such as polymers, fluid mixtures, gels, ferroelectrics, membranes, superfluids, superconductors and the like.

   According to classical theories of phase-ordering dynamics, the domain growth follows a characteristic power-law behavior $L (t) \sim t^\eta$, 
where $L$ is the average domain size and $\eta$ is the scaling exponent. 
The dynamics is universal such that the time evolution of an observable is solely governed by $L(t)$. 
In practice, this scaling hypothesis is tested by the equal-time correlation function 
$C(\br^\prime, t) = \langle \phi(\br + \br^\prime, t)  \phi(\br, t)  \rangle$, 
where $\phi (\br)$ is an order parameter characterizing the dynamical evolution of the system and 
$\langle .. \rangle$ denotes the statistical average. 
The Fourier transform of $C(\br, t)$ is the structure factor  $S(\bk, t) = \langle \phi_{\bk}( t)  \phi_{-\bk}( t)  \rangle$, 
where $\phi_{\bk}$ is the Fourier transform of $\phi (\br)$. 
From a dimensional consideration the structure factor obeys the scaling relation $S(k, t) = L^d f( k L(t) )$ \cite{Bray1994},
%
%
 %
where $d$ is the spatial dimensionality and the scaling function $f(q)$ is independent of time.  
Due to the presence of domain walls, $f(q)$ exhibits a power-law tail  $f(q) \sim q^{-(d+1)}$ at large $q$, 
which is referred to as the Porod law \cite{Debye1957}.
The scaling theory hypothesizes various power laws for domain coarsening, which describe the time dependence of characteristic length scales. 
The domain growth law for a two-dimensional (2D) conserved field is $L \sim t^{1/3}$, 
which characterizes the diffusive transport of the order parameter \cite{Bray1994, Lifshitz1961, Huse1986}. 
In binary fluids a competition between the viscous and inertial flow leads to two growth regimes: 
viscous hydrodynamic  ($L \sim t$) \cite{Siggia1979} and inertial hydrodynamic  ($L \sim t^{2/3}$) \cite{Furukawa1985}.
The viscous hydrodynamic regime has been confirmed by experiments as well as simulations \cite{Bray1994}.
However, the inertial hydrodynamic regime has not been observed yet as viscous flow is non-negligible in classical fluids. 



        Ultracold atoms have emerged as an ideal platform to study the dynamics of multicomponent superfluids,   
forming the basis for the study of a multitude of phenomena such as the miscible-immiscible transition in binary fluids \cite{Mineev, Ho1996,  Timmermans1998, Pu1998}.
Experimentally, Bose-Bose mixtures using different hyperfine levels or different isotopes have been used to study phase separation  \cite{Hall1998, Wieman2008, McCarron2011, Wacker2015}, nonlinear dynamical excitations \cite{Maddaloni2000, Mertes2007, Eto2016, Eto20162}, solitons \cite{Hamner2011, Hoefer2011}, and Townes solitons \cite{Bakkali2021}. 
Domain formation and coarsening were observed in quenched immiscible mixtures \cite{Spielman2014}. 
Following the proposal \cite{Davis2011}, Rabi-coupled Bose mixtures \cite{Nicklas2011} were used to experimentally test the Kibble-Zurek mechanism \cite{Nicklas2015}.
Theoretically, many studies reported dynamical scaling laws in superfluid systems  \cite{ Damle1996, Mukerjee2007, Takeuchi2012, Karl2013, Kudo2013, Karl2013, Kudo2015, Williamson2016, Williamson2016_2, Takeuchi2018, Kazuya2020} and dynamical instabilities \cite{Kasamatsu2004, Kasamatsu2006, Sasaki2009, Takeuchi2010, Suzuki2010,  Sasaki2011, Sasaki2011_2}.
It was pointed out that thermal fluctuations suppress the phase separation at nonzero temperatures \cite{Schaeybroeck2013, Roy2015, Nick2016}. 
The inertial hydrodynamic regime was investigated theoretically in binary 2D superfluids \cite{Kudo2013, Hofmann2014}.   
Nevertheless, many features of dynamical scaling such as the role of thermal fluctuations are yet to be explored.

    In this paper, we investigate how the thermal fluctuations influence the dynamical scaling of coarsening in a binary 2D Bose superfluid. 
To this end, we employ semiclassical-field simulations to address the demixing dynamics at nonzero temperature, which is triggered by quenching the interspecies interaction parameter.       
As a key result, we show how the characteristic scaling of domain growth is modified by temperature and quench parameter. 
The interplay of the quench and thermal energy results in two phase separation regimes. 
The first one is the strong separation regime that occurs at low temperature and high quench, 
for which the Porod scaling law of the structure factor $S(k) \sim k^{-3}$  holds, and 
the domain coarsening follows the inertial hydrodynamic growth law of binary fluids, i.e., $L(t) \sim t^{2/3}$.
The other regime is the weak separation regime at high temperatures, 
where thermal fluctuations modify both the domain growth law and the Porod tail of the structure factor. 
Near the superfluid critical temperature the domain coarsening approaches the diffusive growth law of a 2D conserved field ($L \sim t^{1/3}$) 
and the Porod tail of the structure factor scales close to $S(k) \sim k^{-1}$.  
Furthermore, we examine the demixing dynamics in a box cloud and find an intriguing interplay of the box symmetry and the dynamics. 
In particular, for low quench and small clouds, with sizes comparable to the spin healing length, demixing occurs via the creation of domains of regular patterns due to the boundary condition. 
For high quench and large clouds we recover the dynamics that is similar to the system with periodic boundary conditions.

  



%
\begin{figure}
\includegraphics[width=1.0\linewidth]{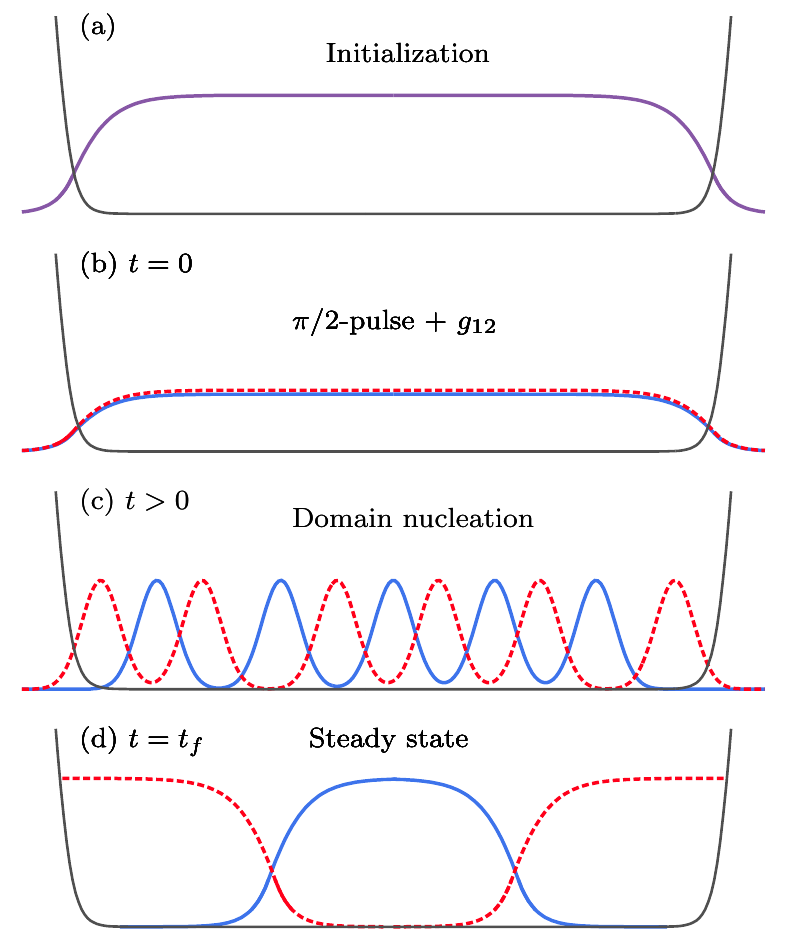}
\caption{Sketch of the quench protocol for a homogeneous cloud confined in a box potential.
(a) The initial state is a thermal state having total density $n$ at temperature $T$.  
(b) We apply a $\pi/2$ pulse to obtain a uniform superposition of species 1 (dashed line) and 2 (continuous line). 
Thereafter, we quench the interspecies interaction $g_{12}$ into the demixed regime, 
which creates a dynamically unstable mixture of two species. 
(c) Time evolution proceeds via nucleation of domains of species 1 and 2. 
(d) Long time evolution at final time $t_f$ results in a steady state having species 2 at the center and species 1 forming a shell around it.
}
\label{Fig:sketch}
\end{figure}

\begin{figure*}
\includegraphics[width=1.0\linewidth]{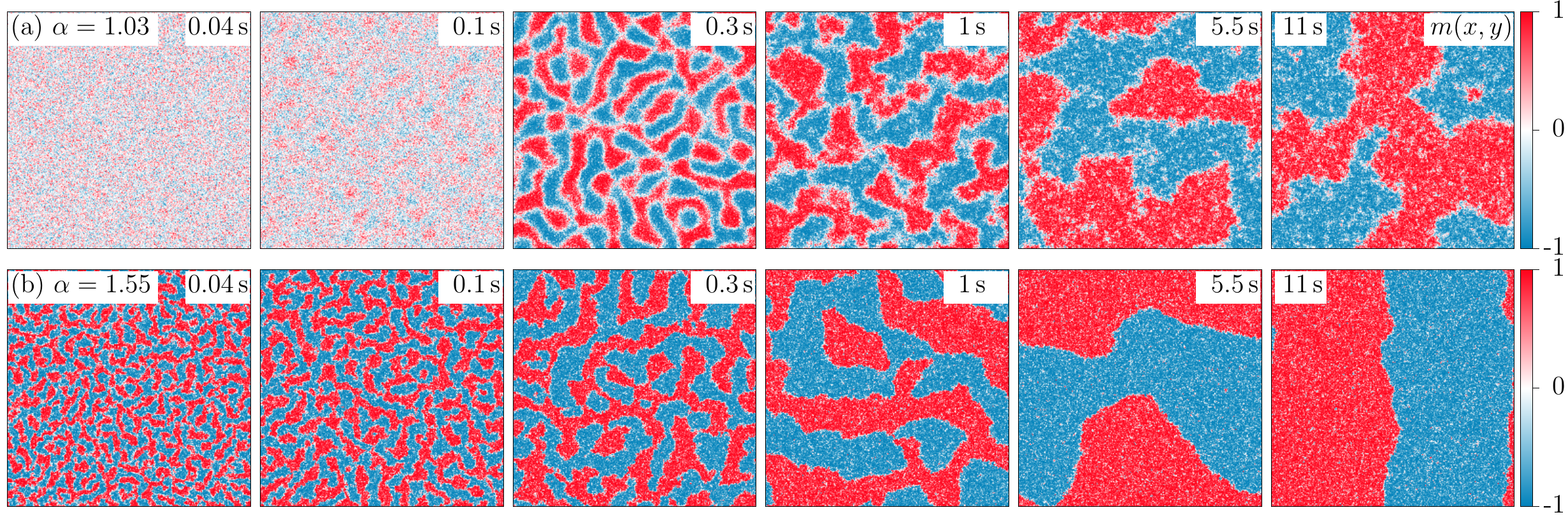}
\caption{Nucleation of domains and the coarsening dynamics. Time evolution of the two-species density imbalance $m(x, y)$ of a single trajectory for  $\alpha =1.03$ (upper row) and $1.55$ (lower row), displaying nucleation of domains of two components (red and blue) and their coarsening dynamics.  
The spatial dimensions for each panel are  $256\times 256\, \mum^2$. }
\label{Fig1}
\end{figure*}

\begin{figure}
\includegraphics[width=1.0\linewidth]{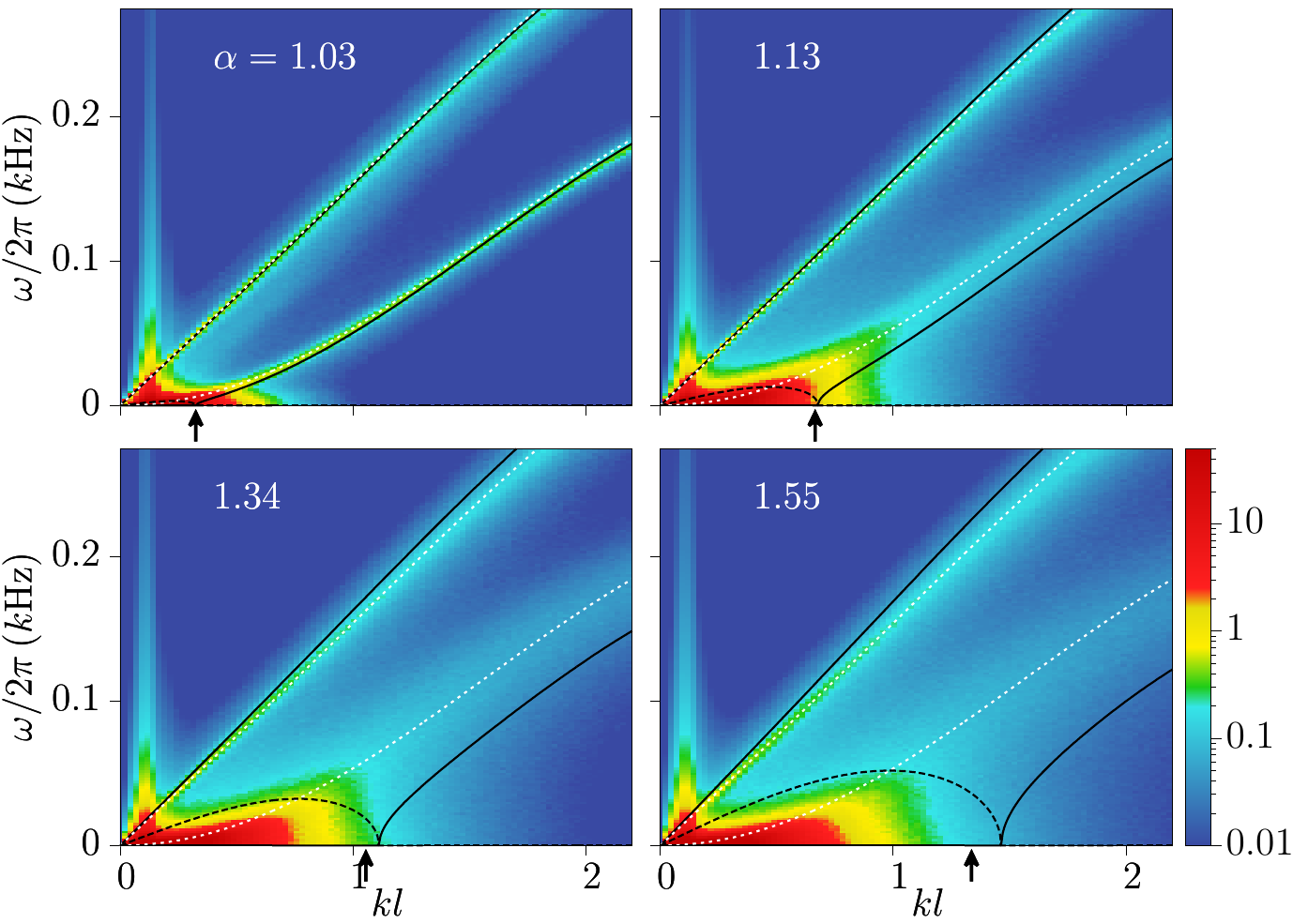}
\caption{Nonequilibrium excitation spectra. Dynamic structure factor $S_1(\bk, \omega)$ as a function of the wave vector $\bk = k \uvec{e}_x$ and frequency $\omega$ for $\alpha=1.03$, $1.13$, $1.34$ and $1.55$. 
The black continuous lines are the Bogoliubov spectra $E_{k, \pm}$ of Eq. \ref{eq_spec} and the black dashed line represents the imaginary part of $E_{k, -}$.
The arrow indicates the momentum range of unstable modes based on Eq. \ref{eq:ku}. 
The white dotted lines correspond to the spectra of phase separated clouds; see text.    }
\label{Fig:dsf}
\end{figure}
\begin{figure*}
\includegraphics[width=1.0\linewidth]{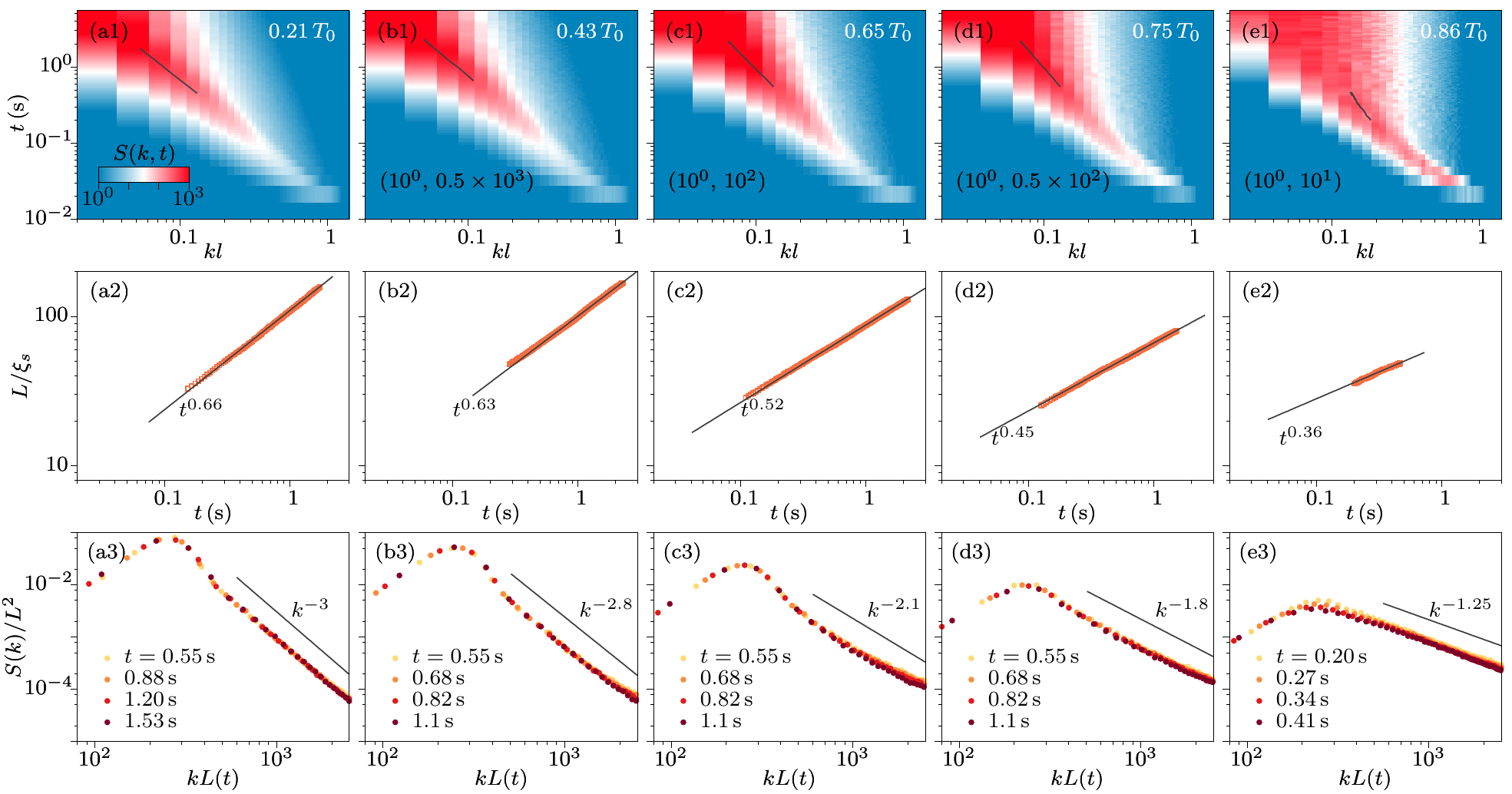}
\caption{Growth laws and dynamical scaling at nonzero temperatures.  
(a1-e1) Structure factor $S(\bk, t)$ as a function of the wave vector $\bk = k \uvec{e}_x$ and time $t$ on a log-log scale for various $T/T_0$. 
The upper and lower range of spectral weights for (b1-e1) are given in parenthesis. 
The location of the maximum (continuous line) allows us to determine the average domain size $L(t)$; see text.
(a2-e2) The values of $L(t)$ and their power-law $t^\eta$ fit (line) yields $\eta= 0.66$, $0.63$, $0.52$, $0.45$ and $0.36$ for $T/T_0=0.21$, $0.43$, $0.65$, $0.75$ and $0.86$, respectively. $\xi_s$ is the spin healing length; see text. 
(a3-e3) Plots of $S(k)/L(t)^2$ versus $kL(t)$ demonstrate universal time evolution for various $t$.  
The continuous line indicates an approximate scaling for the high-momentum tail. 
}
\label{Fig:psd}
\end{figure*}

\section{System and methodology} \label{sec:method}
    We consider a cloud of $^{87}$Rb atoms in two different hyperfine states $|F=1, m_F=0 \rangle$ and $|F=2, m_F=0 \rangle$, which is motivated by the experiments \cite{saintjalm, lecerf}. 
Thus,  the two species have the same masses ($m_1=m_2=m$) and the intraspecies scattering lengths are $a_{11}/a_\mB=100.86 $ and $a_{22}/a_\mB =94.58$ \cite{Widera2006}, where $a_\mB$ is the Bohr radius. 
The interspecies scattering length is $a_{12}/a_\mB=98.9$  \cite{Altin2011}, 
resulting in the parameter $\alpha = 1.012$, which is defined as 
\begin{align}
\alpha \equiv a_{12}/\sqrt{a_{11} a_{22}}.
\end{align}
Since $\alpha$ is slightly above $1$, the two species are weakly immiscible and thermal fluctuations play a prominent role in the demixing dynamics, as we show below. 
We describe the system by the Hamiltonian 
\begin{align} \label{eq_hamil}
\hat{H} =\hat{H}_1 + \hat{H}_2  + \hat{H}_{12},  
\end{align}
with
\begin{align} \label{eq:ha}
\hat{H}_a &= \int d{\bf r} \Big[  \Bigl( \frac{\hbar^2}{2m_a} \nabla \hat{\psi}_a^\dagger({\bf r}) \cdot \nabla \hat{\psi}_a({\bf r})  \nonumber \\ 
& \quad  + \frac{g_{aa}}{2} \hat{\psi}_a^\dagger({\bf r})\hat{\psi}_a^\dagger({\bf r})\hat{\psi}_a({\bf r})\hat{\psi}_a({\bf r}) \Bigr)  \Big], 
\end{align}
and
\begin{align} \label{eq:h12} 
\hat{H}_{12} =   \int d{\bf r} \Big[  g_{12} \hat{\psi}_1^\dagger({\bf r})\hat{\psi}_2^\dagger({\bf r})\hat{\psi}_2({\bf r})\hat{\psi}_1({\bf r})  \Big],
\end{align}
where $a=1, 2$ represent the two species and $\hat{\psi}_{a}$ ($\hat{\psi}_{a}^\dagger$) are the corresponding annihilation (creation) operators.
The intraspecies interactions $g_{aa}$ and interspecies interaction $g_{12}$ are given by, respectively,   
\begin{align}
g_{aa} = \frac{2\sqrt{2 \pi} \hbar^2}{m} \frac{a_{aa}}{\ell_z}  \quad \text{and} \quad g_{12} = \frac{2 \sqrt{2 \pi} \hbar^2}{m} \frac{a_{12}}{\ell_z}.
\end{align}
$\ell_z = \sqrt{\hbar/(m \omega_z)}$ is the harmonic oscillator length of the trapping potential in the transverse direction, 
where $\omega_z$ is the trap frequency. 
For a condensate with a large number of atoms we replace $\hat{\psi}_{a}$ by complex numbers $\psi_{a}$. 
Using Eq. \ref{eq_hamil} we obtain the coupled equations of motion
\begin{align} 
i\hbar \partial_t \psi_1 &= \Bigl(- \frac{\hbar^2}{2m} \nabla^2 + g_{11} n_1 + g_{12} n_2  \Bigr) \psi_1,  \label{eq_eom1} \\
i\hbar \partial_t \psi_2 &= \Bigl(- \frac{\hbar^2}{2m} \nabla^2 + g_{22} n_2 + g_{12} n_1  \Bigr) \psi_2,  \label{eq_eom2}
\end{align}
which govern the dynamics of binary condensates. $n_a= |\psi_a|^2$ are the densities. 
This system hosts two excitation branches of collective modes \cite{Pethick2008}
\begin{equation}\label{eq_spec}
E_{k, \pm}^2 = \frac{(E_{1}^2 + E_{2}^2)}{2} \pm \frac{1 }{2}\sqrt{ (E_{1}^2 - E_{2}^2)^2 + 16 \epsilon_k^2 n_1 n_2 g_{12}^2} , 
\end{equation}
where $E_{a} = \sqrt{ \epsilon_k (\epsilon_k  + 2 g_{aa} n_a)}$ are the single-component Bogoliubov spectra and $\epsilon_k = \hbar^2 k^2/(2m)$.  
The coupling $g_{12}$ results in hybridized branches $E_{k, \pm}$. 
A direct consequence of this hybridization is that the low-momentum part of $E_{k, -}$ vanishes when $\alpha=1$ (or equivalently $g_{12} = \sqrt{g_{11} g_{22}}$) and becomes imaginary for $ \alpha > 1$. 
This leads to the creation of unstable modes when $ \alpha$ is above $1$, which is responsible for the demixing of the two species. 
We note that $\alpha_c = 1$ is the quantum critical point separating the miscible ($\alpha < \alpha_c$ ) and immiscible ($\alpha > \alpha_c$) states at zero temperature \cite{Timmermans1998}. 
The range of unstable modes is determined by setting $E_{k, -}=0$, giving 
\begin{align}\label{eq:ku}
k_{0}^2 = \frac{1}{\xi_1^2 \xi_2^2 } \Bigl[ \sqrt{ (\xi_1^2 - \xi_2^2)^2 + 4 \alpha^2 \xi_1^2  \xi_2^2  } -  (\xi_1^2 + \xi_2^2)       \Bigr],
\end{align}
where $\xi_a = \hbar/\sqrt{2m g_{aa} n_a}$ are the single-component healing lengths. 
$k_0$ vanishes when $\alpha=1$ and increases with increasing $\alpha$ for $\alpha > 1$.  
The wavelength $\lambda_0= 2\pi/k_0$ and the lifetime $\tau= \hbar/E_{k_0, -}$ give an estimate of length and time scale for the emergence of domains.


      We investigate the phase-separation dynamics using the classical-field method of Refs. \cite{Singh2016, Singh2017}. 
For the numerical simulations we discretize the space on a lattice of size $N_x \times N_y$ and a discretization length $l= 1 \, \mum$. 
We note that $l$ is chosen to be smaller than or comparable to the healing length and the thermal de Broglie wavelength \cite{Castin2003}. 
This maps the continuum Hamiltonian on the discrete Bose-Hubbard model, 
which introduces $J=\hbar^2/(2ml^2)$ as the tunneling energy and $U_{aa}=g_{aa} l^{-2}$ and $U_{12}=g_{12} l^{-2}$ as the onsite repulsive interactions. 
We use $\omega_z = 2\pi \times 4.6 \, \mathrm{kHz}$, leading to $U_{11}/J = 0.336$ and $U_{22}/U_{11} = 0.938$ \cite{saintjalm}.  
The quench protocol is described in Fig. \ref{Fig:sketch}. 
We start with a 2D superfluid cloud of total density $n=10\, \mum^{-2}$ at temperature $T$. 
The initial states $\psi_1(\br)$ of this system are sampled in a grand-canonical ensemble of chemical potential $\mu$ and temperature $T$ via a classical Metropolis algorithm \cite{Singh2016}. 
We choose $T$ in a wide range of $T/T_0=0.1 - 1.1$, where $T_0$ is an estimate of the critical temperature for the superfluid transition in weakly interacting 2D Bose gases \cite{Prokofev2001, Prokofev2002}. 
For the other species we sample the initial states with vacuum fluctuations, i.e., $\langle |\psi_2(\br_i)|^2  \rangle = 1/(2l^2)$ \cite{LM2017}, where the index $i$ corresponds to the lattice site and $\langle .. \rangle$  denotes the ensemble average.
At time $t=0$ we use a $\pi/2$ pulse to obtain a uniform superposition of the states 
$\psi_{1/2} (\br_i) = [\psi_{1} (\br_i)   \pm  \psi_{2} (\br_i)   ]/ \sqrt{2}$. 
This results in the two cloud densities $n_1 \approx n_2 \approx n/2 = 5\, \mum^{-2}$, since $g_{11}$ and $g_{22}$ are similar.
We then quench $g_{12}$ in the demixed regime ($\alpha >1$) and determine the time evolution $\psi_{1/2} (\br, t)$ via Eqs. \eqref{eq_eom1} and \eqref{eq_eom2}. 
As schematically shown in Figs. \ref{Fig:sketch}(c) and (d), the initial time evolution proceeds via nucleation of small-sized domains and  the long-time evolution results in two spatially separated clouds.
For our simulations we vary $\alpha$ in the range $1.03 \leq \alpha \leq 1.55$ to explore both weakly and strongly immiscible regimes,  
which covers a wide range of immiscible regime that can be realized with mixtures of other species. 
To analyze the demixing dynamics, as an order parameter, we calculate the local density imbalance 
\begin{align}
m(\br, t) = \frac{ n_1(\br, t)  - n_2 (\br, t)}{n_1(\br, t)+n_2(\br, t)}.
\end{align}
We show $m(\br, t)$ for a periodic-boundary system in Fig. \ref{Fig1} and its average $\langle m(\br, t) \rangle$ for a box system in Fig. \ref{Fig:box}, where $\langle .. \rangle$ denotes an average over the initial ensemble.

 \section{Results} 
 \subsection{Demixing dynamics} 

  In Fig. \ref{Fig1} we show the time evolution of $m(\br, t)$ of a single trajectory at $T/T_0=0.21$ for $\alpha=1.03$ and $1.55$.  
We employ a fixed system size of $256\times 256\, \mum^2$ for all periodic-boundary simulations.  
The quench to the demixed state at $t=0$ triggers the nucleation of domains of each component, 
where smaller domains are present for $\alpha=1.55$ than that for $\alpha=1.03$.  
We estimate the initial domain size $L_0$ by the momentum range $k_0$ of unstable modes in Eq. \ref{eq:ku}.  
We obtain $L_0 \sim \lambda_0= 19$ and $4.7\, \mum$ and the nucleation time $t \sim \tau= 26\, \mms$ and $1.5\, \mms$ for $\alpha=1.03$ and $1.55$, respectively.
The intermediate time evolution manifests the coarsening process, where small domains shrink and large ones grow.  
There are small patches of other component in the domains, which are due to the initial fluctuations that suppress the dynamics in the weak separation regime.
This is the scenario for $\alpha=1.03$, whereas for $\alpha=1.55$ the dynamics is weakly affected by these fluctuations as the system is in the strong separation regime. The weak versus strong separation regime occurs as an interplay between the thermal and quench energy. 
At high temperatures thermal fluctuations dominate the dynamics and no phase separation occurs as we show in Appendix \ref{ap:sec:temp}.

    To identify the interplay of collective modes in the demixing dynamics we calculate the dynamic structure factor of the density
 \begin{align}
 S_1(\bk, \omega) = \langle  |n_1(\bk,  \omega) |^2  \rangle,
 \end{align}
where $n_1(\bk,  \omega)$ is the Fourier transform of the density $n_1(\br, t)$ of component 1 in space and time:   
\begin{align}
n_1(\bk,  \omega)  = \frac{1}{\sqrt{N_l  T_s}  } \sum_j \int dt\,  e^{- i (\bk \br_j -\omega t ) } n_1(\br_j, t). 
\end{align}
$T_s=0.55\, \ms$ is the sampling time for the numerical Fourier transform and $N_l=N_x N_y$ is the number of lattice sites. 
In Fig. \ref{Fig:dsf} we show $S_1(\bk, \omega)$ as a function of the wave vector $\bk = k \uvec{e}_x$ and frequency $\omega$ for $\alpha=1.03$, $1.13$, $1.34$ and $1.55$.  
We observe both dynamically stable and unstable modes, 
where stable modes appear as two excitation branches and unstable ones as a broad spectrum of low-energy excitation. 
We compare these results with the Bogoliubov spectra $E_{k, \pm}$ of Eq. \ref{eq_spec}. 
For our discretized system, the free-particle dispersion takes the form $\epsilon_k = 2J[1 - \cos(k_xl)]$, 
where $J= \hbar^2/(2ml^2)$ is the tunneling energy. 
We show the real-valued predictions of $E_{k, \pm}$ as the continuous lines in Fig. \ref{Fig:dsf}, 
which capture the excitation branches of stable modes for low and intermediate $\alpha$ and show deviations for high $\alpha$. 
We also show the imaginary solution of $E_{k, -}$, which qualitatively captures the broad spectrum of unstable modes. 
The momentum range of unstable modes increases with increasing $\alpha$ and is close to the predictions of Eq. \ref{eq:ku}.
There is a peak-like excitation at small $k$ corresponding to the structure of a macroscopic domain that the system forms at time $t=T_s$. 
This excitation peak shifts to a lower $k$ for large $\alpha$, implying a rapid growth of domains at large $\alpha$, 
which is consistent with the dynamics shown in Fig. \ref{Fig1}.

   Furthermore, we compare the two excitation branches of stable modes with the spectra of phase separated clouds.  
In this case,  the cloud density is twice the initial density, i.e.,  $n_{i, f} = 2n_i$ and the Bogoliubov spectrum reads  $E_{k, n_{1, f} } = \sqrt{\epsilon_k (  \epsilon_k + 2 g_{11} n_{1, f}  )}$. 
This result agrees with the upper branch of $S_1(k, \omega)$ for all $\alpha$ in Fig. \ref{Fig:dsf}.
The other component being spatially separated from component 1 acts as a thermal cloud 
whose free-particle dispersion captures the lower branch for all $\alpha$ in Fig. \ref{Fig:dsf}.

\begin{figure}
\includegraphics[width=1.0\linewidth]{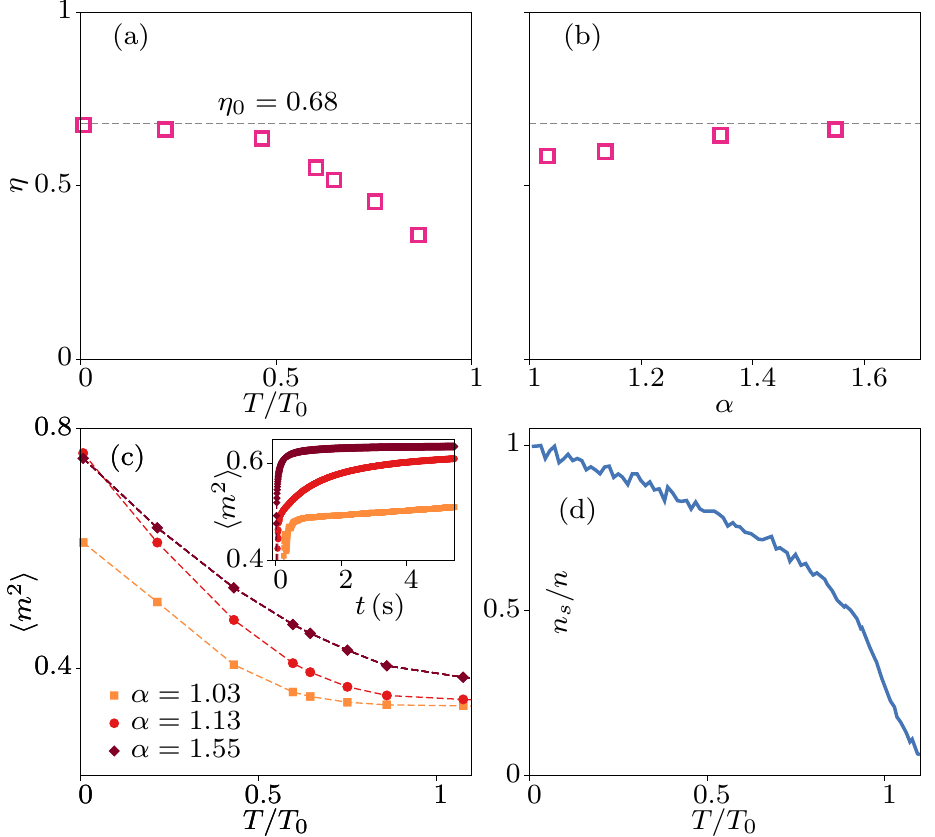}
\caption{Temperature and quench-parameter dependence.  
(a) $\eta(T)$ for $\alpha=1.55$. (b) $\eta(\alpha)$ at $T/T_0=0.21$. 
The horizontal dashed line marks the zero-temperature prediction $\eta_0=0.68$.  
(c)  Average squared imbalance $\langle m^2 \rangle$ at time $t=5.5\, \ms$ as a function of $T/T_0$ for $\alpha=1.03$, $1.13$ and $1.55$, 
while inset shows the time evolution  at $T/T_0=0.21$. 
(d) Temperature dependence of the initial superfluid fraction $n_s/n$, which we determine using the method described in Ref. \cite{Singh2021}.
The results are obtained for the system size $256\times 256\, \mum^2$. 
  }
\label{Fig:eta}
\end{figure}

\begin{figure*}[t]
\includegraphics[width=1.0\linewidth]{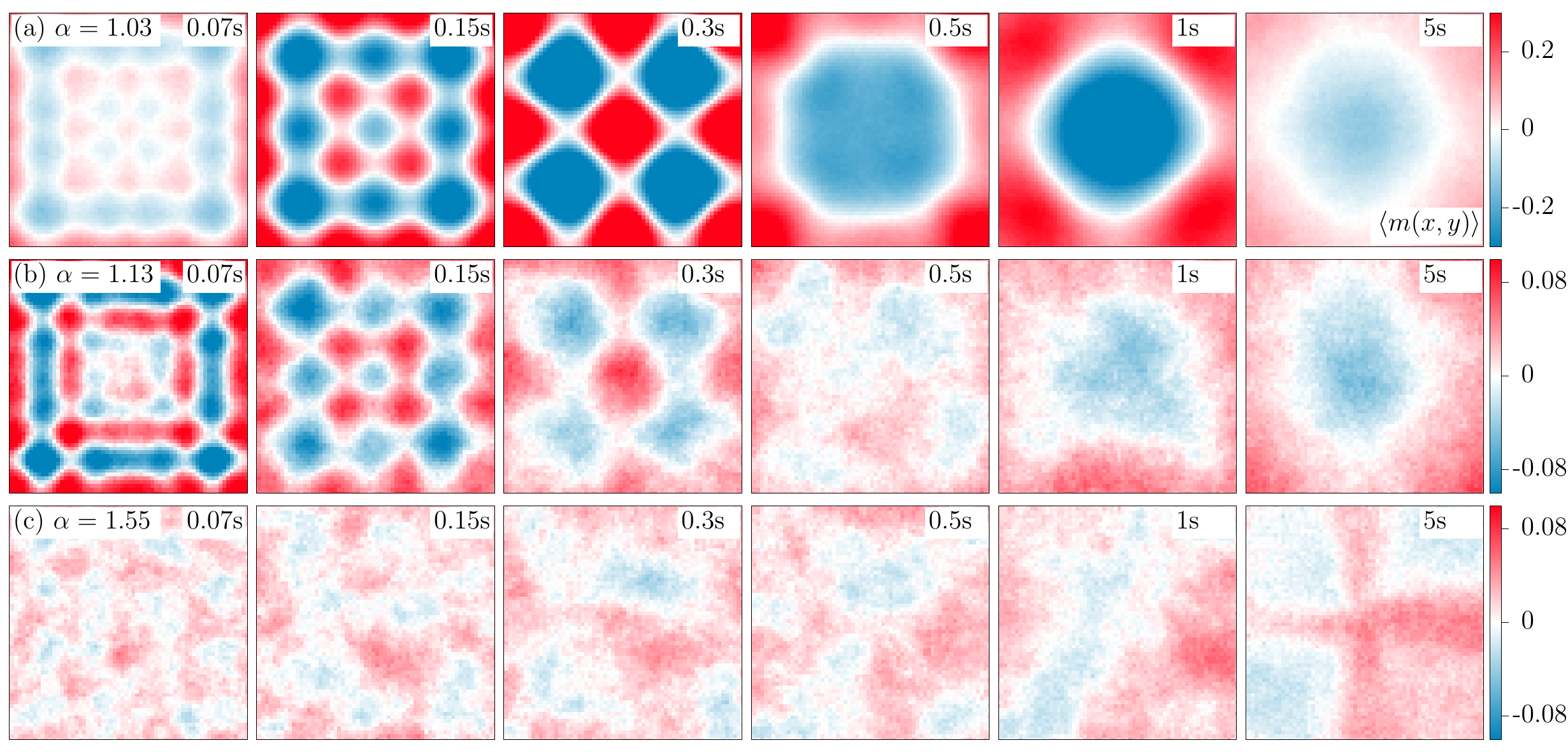}
\caption{Domain formation in a square box of size $64\times 64\, \mum^2$. 
(a-c) show the time evolution of the average imbalance $\langle m(x, y) \rangle$ at $T/T_0= 0.21$ for $\alpha =1.03$ (upper row), $1.13$ (middle row) and $1.55$ (lower row).  
We present the full time evolution as videos in the Supplementary material, which displays a continuous transformation between the domains of different shapes.   }
\label{Fig:box}
\end{figure*}

 \subsection{Dynamical scaling} 

  To characterize the scaling behavior we calculate the structure factor of the imbalance 
 \begin{align}
 S(\bk) = \langle  |m(\bk) |^2  \rangle,
 \end{align}
with
\begin{align}
m(\bk) = \frac{1}{\sqrt{N_l} } \sum_j \exp(- i \bk \br_j ) m(\br_j),
\end{align}
where $m(\bk)$ is the Fourier transform of $m(\br)$. 
In Figs. \ref{Fig:psd}(a1-e1) we show $S(\bk, t)$ as a function of the wave vector $\bk = k \uvec{e}_x$ and time $t$ for $\alpha=1.55$ and  various $T/T_0$. 
The nucleation of domains is indicated by the spectral peak at finite $k$, which gradually moves to smaller $k$ as the domains coarsen. 
The location of the peak describes an average size of the domain, whereas the peak broadening reflects the influence of thermal fluctuations on the dynamics. 
The thermal effect is strong at high temperature, resulting in a decreasing peak amplitude in Figs. \ref{Fig:psd}(a1-e1). 
For $T/T_0= 0.86$, only after a short time evolution, the spectral peak vanishes due to strong diffusion induced by thermal fluctuations.
We fit the structure factor with the Gaussian distribution $g(k) = A_0 \exp[ -(k-k_d)^2/(2 \sigma^2) ]$, 
where $A_0$, $k_d$, and $\sigma$ are the fitting parameters. 
From $k_d$ we determine the average domain size $L = 2\pi/k_d$. 
In Figs. \ref{Fig:psd}(a2-e2) we show the determined values of $L(t)$ on a log-log scale. 
The growth of $L(t)$ demonstrates a power-law behavior that is typical for coarsening of macroscopic domains $L \gg \xi_s$, 
where  the spin healing length is defined as $\xi_s = \hbar/\sqrt{2m n g_s}$, with $g_s=(2g_{12} - g_{11} - g_{22})/2$.  
$\xi_s$ is a length scale on which the two species interact to nucleate domains.  
We fit $L(t)$ with the function $f(t) = c_0  t^\eta$, where $c_0$ is the fitting parameter. 
This way, we determine the scaling exponent $\eta$,  see caption of Fig. \ref{Fig:psd}.
We note that the value of $\eta$ decreases with increasing temperature.

  In Figs. \ref{Fig:psd}(a3-e3)  we show the scaled structure factor $S(k, t)/L(t)^2$ as a function of the scaled wave vector $k L(t)$ for various $t$ and $T/T_0$.
The different-time results collapse on one single curve, confirming the scaling hypothesis.
The momentum tail follows a power-law behavior and its decay varies with temperature. 
For $T/T_0=0.21$ we find a dependence $S(k) \sim k^{-3}$, which is consistent with the Porod law \cite{Bray1994}.  
This occurs due to the presence of domain walls that lead to the dependence $S(k) \sim k^{-3}$ at large $k$ in 2D. 
At high temperature, the momentum tail decays slowly as the process of phase separation is suppressed by dominant thermal fluctuations. 
For $T/T_0 = 0.86$ we find a momentum-tail behavior $S(k) \sim k^{-1.25}$, where no phase separation is visible in the time evolution, 
see Appendix  \ref{ap:sec:temp}.


%
\begin{figure}[t]
\includegraphics[width=1.0\linewidth]{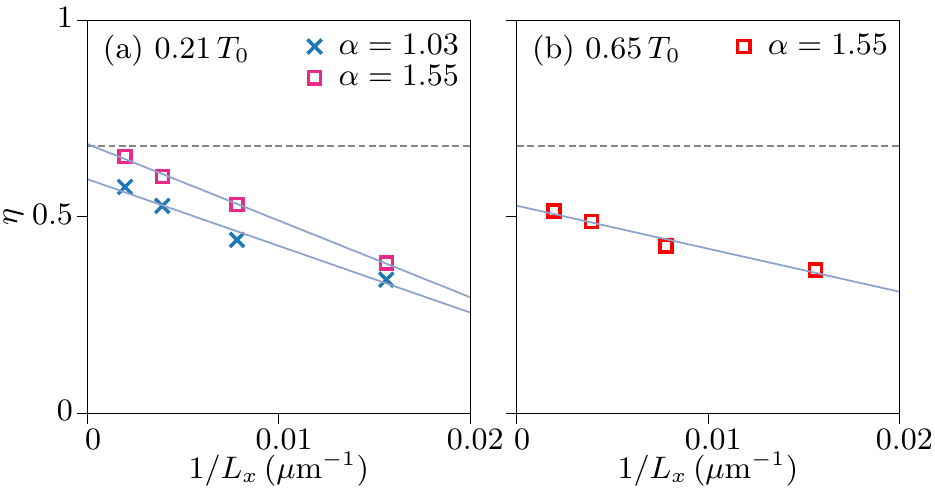}
\caption{Finite-size scaling.  
(a) $\eta$ as a function of the inverse system length $1/L_x$ at  $T/T_0= 0.21$ for $\alpha=1.03$ (crosses) and $1.55$ (squares).  
The linear fit (continuous lines) yields  $\eta_ \infty =0.59$ and $0.68$ for $\alpha = 1.03$ and  $1.55$, respectively.  
(b) We obtain $\eta_ \infty  = 0.53$ at  $T/T_0= 0.65$ for $\alpha = 1.55$.   }
\label{Fig:size}
\end{figure}

        Figs. \ref{Fig:eta}(a) and (b) show the temperature and quench-parameter dependence of the scaling exponent. 
Both $T$ and $\alpha$ influence the value of $\eta$, exemplifying the interplay between strong and weak separation regimes. 
At low $T$ and large $\alpha$, the system is in the strong separation regime in the sense that the dynamics is marginally affected by initial fluctuations. 
Here we obtain $\eta$ close to the zero-temperature prediction $\eta_0= 0.68$, which suggests that the dominant process for domain growth is the inertial hydrodynamic transport of superfluid from low-density to high-density regions. 
This scenario is consistent with the simulations of binary condensates at zero temperature \cite{Hofmann2014}.
At high $T$ and small $\alpha$, initial fluctuations suppress the dynamics of phase separation and result in a  renormalization of $\eta$ and the high-momentum tail of the structure factor. We refer to this regime as the weak separation regime. 
To quantify these regimes we calculate the time evolution of the average squared imbalance $\langle m^2 \rangle$. 
In Fig. \ref{Fig:eta}(c) we show $\langle m^2 \rangle$ at $t=5.5\, \ms$ as a function of $T/T_0$ for $\alpha = 1.03$, $1.13$ and $1.55$.  
It decreases with increasing $T/T_0$ and decreasing $\alpha$. 
As shown in the inset of Fig. \ref{Fig:eta}(c), $\langle m^2 \rangle$ increases during the time evolution and reaches the steady state in the long-time evolution. 
We find that the dynamics is in the strong separation regime for $\langle m^2 \rangle \gtrsim 0.64$, 
where we recover both the zero-temperature $\eta_0$ and the Porod tail of the structure factor.
The weak separation regime sets in when $\langle m^2 \rangle \lesssim 0.64$, where thermal fluctuations renormalize the scaling parameters.
Here the dynamics is influenced by thermal fluctuations that suppress the initial superfluid order of the system [Fig. \ref{Fig:eta}(d)]. 
For  $\langle m^2 \rangle \lesssim 0.4$ we observe no phase separation. 

 \subsection{Demixing dynamics in a square box} 

    We now turn to the demixing dynamics of a homogeneous 2D cloud confined in a square-box geometry,  
which is motivated by the experiments \cite{saintjalm, lecerf}.  
Compared to a periodic boundary system, where domain locations are spontaneous, 
finite boundaries break the translational invariance and act as a pinning potential for the formation of domains \cite{Spielman2014}. 
We choose the same density and the same quench protocol as above.
We first analyze the demixing dynamics in a box cloud of size $64\times 64\, \mum^2$, 
which is comparable to the experiments  \cite{saintjalm, lecerf}.
In Fig. \ref{Fig:box} we show the time evolution of the average imbalance $\langle m(x, y) \rangle$ at $T/T_0=0.21$ 
for $\alpha=1.03$, $1.13$, and $1.55$.  
Indeed, the nucleation of domains is pinned by the box boundaries, which stems from a density difference at the edges since $a_{11}$ and $a_{22}$ are different,  serving as a seed for the creation of domains. 
On the contrary, in the case of periodic-boundary systems domain nucleation is seeded from the fluctuations of the field. 
The box symmetry results in a qualitatively different average dynamics than in infinite systems. 
For $\alpha=1.03$, the time evolution proceeds via formation of regular patterns that undergo a continuous transformation to create structures of striking, geometric shape. 
At $t \sim 0.5\,  \ms$ the time evolution shows the creation of one macroscopic domain of component 1 and 2, 
which then equilibrates after $t \sim 5 \, \ms$.
These results are close to the measurements that show the creation of similar structures for time evolution up to $100\, \mms$  \cite{saintjalm, lecerf}. 
The pinning effect is suppressed when $\alpha$ is high, see the dynamics for $\alpha=1.13$ and $1.55$ in Figs. \ref{Fig:box}(b) and (c).  
The reason for this is the smaller spin healing length $\xi_s$ at higher $\alpha$, which supports the creation of small-sized domains.  
 $\xi_s$ is $4.6$, $2.2$ and $1.1\, \mum$ for $\alpha=1.03$, $1.13$, and $1.55$, respectively. 
 For the cloud size considered, we obtain $L_x/\xi_s \approx 14$ and $58$ for $\alpha=1.03$ and $1.55$, respectively, 
 where the former supports the creation of regular-shaped domains due to the boundary condition. 
So, for $L_x/\xi_s \gg 1$, the dynamics approaches the one obtained for a system with periodic boundary conditions.




        Next, we analyze the growth laws for the domains in a box cloud of sizes between $64\times 64\, \mum^2$ and $512\times 512\, \mum^2$.  
 We calculate the structure factor $S(k, t)$ to determine the average domain size using the procedure described above. 
From the power-law growth of domains we ascertain the scaling exponent $\eta$, which is analogous to Fig. \ref{Fig:psd}.  
In Fig. \ref{Fig:size}(a) we show $\eta$ as a function of $1/L_x$ at $T/T_0=0.21$ for $\alpha=1.03$ and $1.55$.  
$L_x$ is the linear dimension of the box. 
The variation of system size by a factor of $64$ allows us to perform a reliable finite-size scaling,  
which gives access to the scaling exponent  $\eta_ \infty$ in the thermodynamic limit. 
We obtain $\eta_ \infty =0.59$ and $0.68$ for $\alpha = 1.03$ and  $1.55$, respectively.  
The results of $\eta_ \infty$ are close to the values obtained for periodic-boundary conditions. 
In Fig. \ref{Fig:size}(b) we show $\eta$ as a function of $1/L_x$  at $T/T_0=0.65$ for $\alpha=1.55$. 
For this system we find $\eta_ \infty  = 0.53$, confirming the renormalization of the scaling exponent at high temperature.


\section{Conclusion}

   We have studied the demixing dynamics of a binary 2D Bose superfluid using classical-field simulations. 
By quenching the interspecies interaction parameter  
we have analyzed the coarsening dynamics at various values of temperature and the quench parameter. 
We have demonstrated that the dynamical scaling of domain growth interpolates between the inertial hydrodynamic growth law of binary fluids and the diffusive growth law of a 2D conserved field. 
Specifically, for low temperature and high quench we have found the inertial hydrodynamic growth law $L(t) \sim t^{2/3}$ and the Porod scaling law of the structure factor $S(k) \sim k^{-3}$, where $L$ is the average domain size and $k$ is the wave vector.
We have pointed out that at high temperature thermal fluctuations suppress the demixing dynamics and 
modify both the domain growth law and the Porod tail of the structure factor. 
We have shown that near the superfluid transition temperature the scaling dynamics approaches the diffusive growth law of a 2D conserved field, 
 $L(t) \sim t^{1/3}$, and the Porod tail scales similar to $S(k) \sim k^{-1}$. 
We have then studied the demixing dynamics in a box cloud. 
We have shown that for low quench and small clouds of sizes comparable to the spin healing length  
the box symmetry gives rise to distinctive dynamics, which is characterized by domains of geometric shapes.  
By varying the system size we have determined the scaling exponents of the growth law and found them to be consistent with the results of systems with periodic boundary conditions.

 Our results highlight the fundamental interplay of the quench and thermal energy in phase separation, 
which modifies the underlying scaling laws of coarsening dynamics.  
The experimental realization of our results provides a quantum simulation of scaling laws of binary fluids.
Furthermore, Bose mixtures in a ring trap offer the capability to study the solid-body rotation and persistent currents in multicomponent quantum mixtures.

\section*{Acknowledgements} 
We thank Jean Dalibard and  Rapha{\"e}l Saint-Jalm for inspiring discussions.  
L. M. acknowledges funding by the Deutsche Forschungsgemeinschaft (DFG) in the framework of SFB 925 – project ID 170620586 and the excellence cluster  `Advanced Imaging of Matter’ - EXC 2056 - project ID 390715994.

\begin{figure*}
\includegraphics[width=1.0\linewidth]{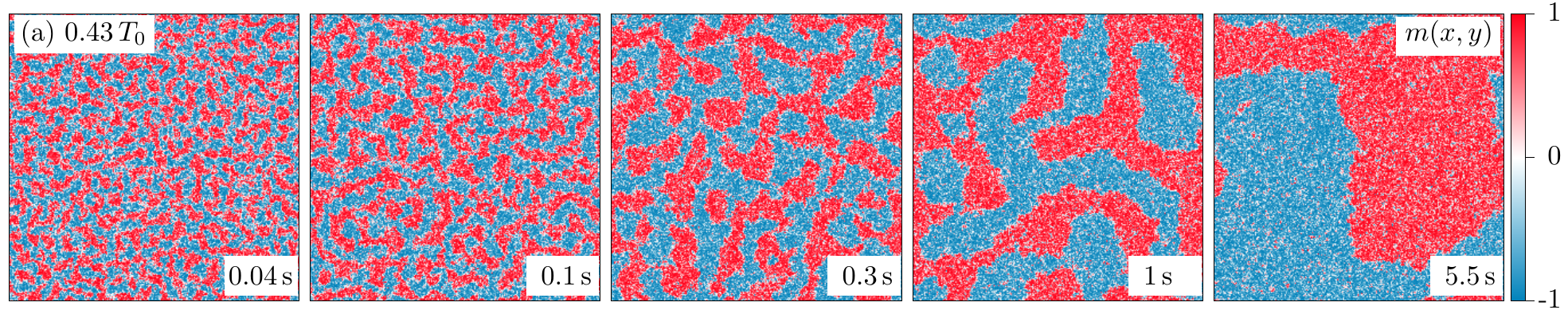}
\includegraphics[width=1.0\linewidth]{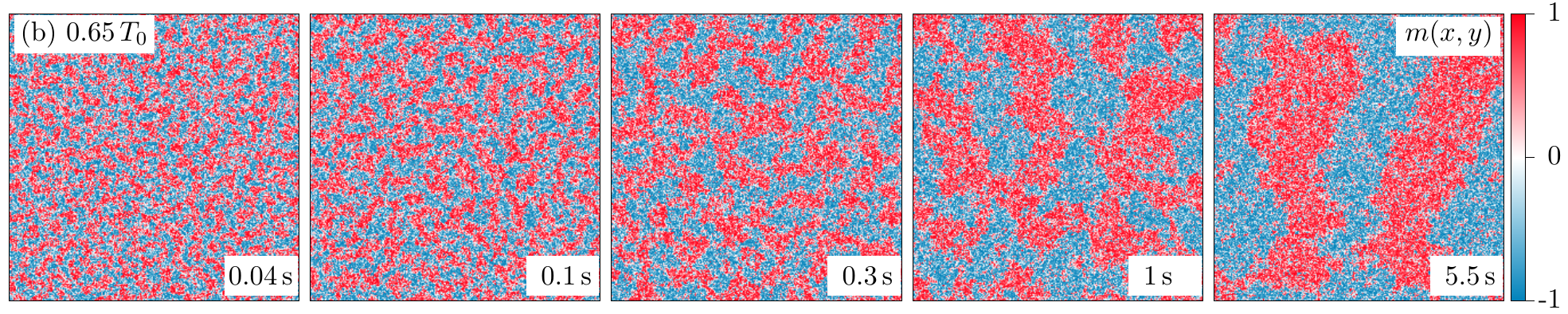}
\includegraphics[width=1.0\linewidth]{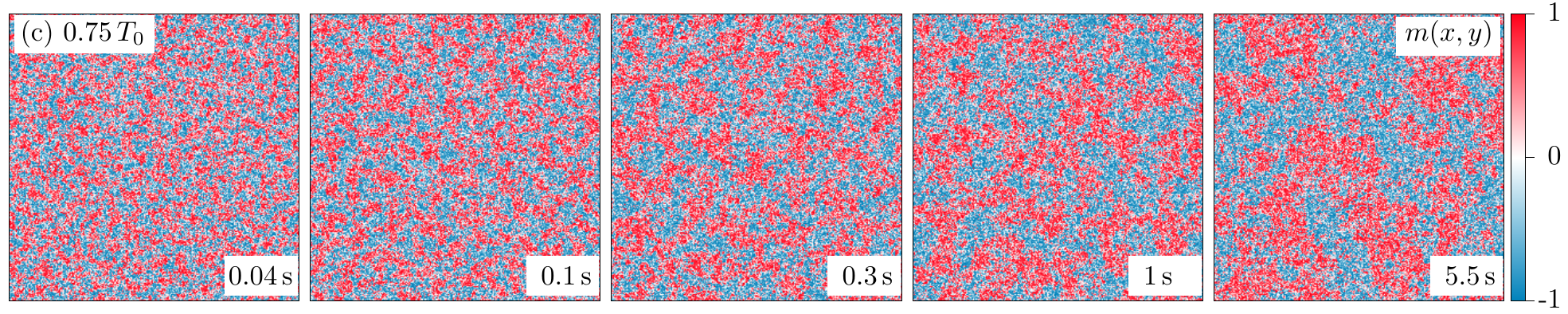}
\includegraphics[width=1.0\linewidth]{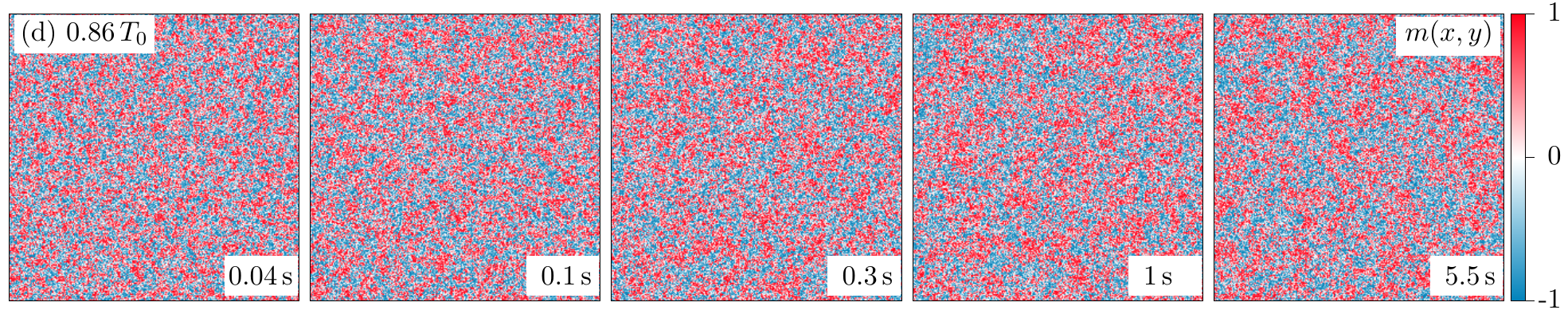}
\caption{Influence of temperature on the demixing dynamics in a periodic-boundary system of size $256\times 256\, \mum^2$ for $\alpha=1.55$. 
(a-d) show the time evolution of the imbalance $m(x, y)$ of a single trajectory after the quench deep in the demixed state at temperatures $T/T_0= 0.43$, $0.65$, $0.75$ and $0.86$. 
The red and blue colors denote the two components. }
\label{FigA1}
\end{figure*}

\appendix
  
  \section{Influence of temperature on the demixing dynamics}\label{ap:sec:temp}

    In this section we show how the thermal fluctuations suppress the demixing dynamics at nonzero temperatures. 
We calculate the imbalance $m(x, y)$ of a single sample of the ensemble for the miscible parameter $\alpha =1.55$ 
and the system size $256 \times 256 \, \mum^2$.   
In Fig.  \ref{FigA1} we show the time evolution of  $m(x, y)$ for various values of the temperature $T/T_0$. 
The coarsening dynamics is affected by the initial thermal fluctuations, resulting a suppression of phase separation at high temperatures. 
Near the superfluid critical temperature at $T/T_0= 0.86$ the phase separation is indistinguishable.

\bibliography{References}












\end{document}